\documentstyle[psfig]{mn}

\begin{document}
\title[The Faint 12$\mu$m Population]{The milliJansky 12$\mu$m Population:
First Followup \thanks{Based in part on observations carried out at
the European Southern Observatory, La Silla, Chile, and in part on
observations with ISO, an ESA project with instruments funded by ESA
Member States (especially the PI countries: France, Germany, the
Netherlands and the United Kingdom) and with the participation of ISAS
and NASA}}
\author[D.L. Clements et al.]
{D.L. Clements$^{1,2}$, F-X. Desert$^{3}$, A. Franceschini$^{4}$\\
$^1$Department of Physics and Astronomy, University of Wales Cardiff,
PO Box 913, Cardiff, CF24 3YB\\ $^2$Institut d'Astrophysique Spatiale,
B\^atiment 121, Universite Paris XI, F-91405 ORSAY CEDEX, France\\
$^3$Observatoire de Grenoble, B.P. 53, 414 Rue de la Piscine, 38041
Grenoble Cedex 9, France\\ $^4$Osservatorio di Padova, Vicolo
Osservatorio, 5, 35122 Padova, Italy}

\maketitle

\begin{abstract}
We present the first results of our followup programme of optical
imaging and spectroscopy of a deep 12$\mu$m survey conducted with the
ISO satellite (Clements et al. 1999). We find that the objects are
typically of fairly low redshift (z$\sim$0.1--0.2), but with a tail
that extends to high redshifts. The highest redshift object is a
previously unknown quasar at z=1.2.  The sample of objects for which
spectroscopy has been obtained forms a complete subsample of the
original survey with an R band magnitude limit of 19.6.  We are thus
able to use {\em accessible volume} methods to determine the
luminosity function.  We find that the luminosity function is close to
the low redshift 12$\mu$m luminosity function determined by IRAS (Fang
et al., 1998; Rush et al., 1993), but with a possible excess at the
highest luminosities, which also correspond to the highest
redshifts. This excess is compatible with the rapid luminosity
evolution of $L \propto (1+z)^{4.5}$ claimed by Xu (2000) and
suggested by Elbaz (1999) on the basis of number counts. These
conclusions remain tentative because of small number statistics at the
highest redshifts.

Almost all of the galaxies in this survey are
emission line objects. We use the spectra to determine emission line
diagnostics for the underlying ionisation source. We find that the
majority of objects have HII region--like spectra (roughly 2/3) but
there is a significant fraction (about 1/3) that contain an
an AGN. This confirms suspicions that deep mid-IR surveys will be a
powerful way to find obscured AGN. Part of the survey area has been
imaged deeply by the INT Wide Field Camera. Nearly all 12$\mu$m sources in
this area, down to the 3$\sigma$ sensitivity limit of the original ISO
survey, have good optical identifications, the faintest being at
R=23. The prospects are thus good for obtaining complete spectroscopy
for this sample. With such data we should be able to make a clear
measurement of the evolution rate of this population and determine the
role of obscured AGN in the mid-IR.
\end{abstract}


\section{Introduction}

The ISO satellite (Kessler et al., 1996) has vastly increased our
observational capabilities in the mid- and far-IR, and has ushered in
a new series of deep cosmological surveys at these wavelengths
(eg. Oliver et al. 2000; Elbaz et al. 1999; Kawara et al. 1998; Puget
et al., 1999; Clements et al. 1999). In the mid-IR these surveys reach
flux limits more than 100 times fainter than our previous best surveys
with IRAS (Hacking et al., 1987), whilst also attaining substantially
improved spatial resolution. They thus represent an important new tool
for observational cosmology. They are especially important since there
is increasing evidence that much of the star-formation activity in the
universe is obscured by dust, which re-radiates the energy absorbed at
optical and UV wavelengths into the mid- and far-IR. While the
discovery of luminous and ultraluminous infrared galaxies (LIGs and
ULIRGs) in the local universe by IRAS was the first hint of the
potential importance of obscured emission (see eg. Saunders \&
Mirabel, 1996, and references therein), it wasn't until the discovery
of the cosmic infrared background (CIB) (Puget et al., 1996) and large
numbers of discrete sources in SCUBA submm surveys (eg. Eales et al.,
1999) and ISOPHOT far-IR surveys (eg. Puget et al., 1999) that the
cosmological significance became clear.  It now seems that such
obscured objects might contribute up to 2/3 of the integrated emission
in the universe (Gispert et al., 2000) and might thus be the sites for
the majority of star formation. Alternatively, a significant number of
the obscured sources might be powered by active nuclei, and the
sources uncovered by mid and far-IR surveys might thus tell us much
about the formation of quasars, and thence the cosmic X-ray background
(Almaini et al., 1999).

While much of the emission from obscured sources will be emitted in
the far-IR, at wavelengths $>$50$\mu$m, there are substantial
advantages in pursuing surveys in the mid-IR, at wavelengths between
10 and 20$\mu$m. These are primarily to do with the limitations of the
current generation of instruments. Since ISO had only a 60cm mirror,
the angular resolution attainable at long, far-IR, wavelengths is
quite poor, leading to problems with confusion and with
identifications for followup surveys. These difficulties are largely
avoided in the mid-IR The mid-IR region was also the most sensitive on
ISO, with improvements up to 100x over IRAS, while the far-IR
typically only reached 10x greater depths, partly as a result of
confusion noise (Puget et al., 1999). Despite operating at shorter
wavelengths, the mid-IR is still capable of detecting
obscured star formation regions. For example, the Antennae
galaxy has optically prominent galactic nuclei, but mid-IR images,
obtained with ISOCAM (Mirabel et al., 1998), show that the bulk of the
star formation resides between these nuclei, in a region of high
obscuration. The peak of the mid-IR flux also coincides with the peak
of both the 850$\mu$m emission, as seen by SCUBA, and the peak of the
molecular emission, as revealed by CO observations (Haas et al., 2000,
and papers in Hughes \& Lowenthal, 2000).

Mid-IR observations, however, do have some problems. These are
principally the result of the complex spectral energy distributions
(SEDs) in the mid-IR band. For starburst galaxies, which are likely to
be the majority of the objects found in a deep mid-IR survey, the SED
is dominated by the so-called Unidentified Infrared Bands (UIB), which
produce prominent broad features at wavelengths of 6.2, 7.7, 8.6, 11.3
and 12.7 $\mu$m, and can vary in strength from object to
object. Furthermore, there is also the possibility of emission and
absorption line contributions, the strongest of which will be the Si
absorption feature at 9.7$\mu$m. The effect of these features on
cosmological surveys in the mid-IR has been extensively studied by Xu
et al. (1998). They find that the effects of these features will be
more pronounced the narrower the observational passband. While most
deep cosmological mid-IR surveys have been conducted in the 15$\mu$m
band, the present work uses the broader 12$\mu$m filter. The Xu et
al. modelling clearly shows that this filter will be less susceptible
to the effects of the complex SED. 

The current results of deep mid-IR surveys with ISO are summarised by
Elbaz et al. (1999; but see also Oliver et al., 2000; Serjeant et
al., 1997; Clements et al., 1999; Serjeant et al., 2000). Substantial evolution in the
number counts, beyond no-evolution extrapolations from the local
population, has been found (eg. Elbaz et al., 1999). The meaning of these
counts, though, is still unclear. Elbaz et al. (1998), for example,
claim that a new population of objects becomes significant at flux
levels of $<$ 1mJy at 15$\mu$m, producing a distinct bump in the
number counts. They suggest that this is a new population of objects
at z$<$1.5 contributing a substantial amount of the star formation
history of the universe.  Meanwhile, Xu (2000) suggests that a
combination of simple luminosity evolution at a rate of L $\propto
(1+z)^{4.5}$ combined with the colour effects of the UIB features can
produce the jump in number counts seen in the data. It is not possible
to decide between these possibilities using the existing number count
data. The only way to determine the true nature and significance of
the sources uncovered by deep mid-IR surveys is to determine their
redshifts, and thus to explicitly measure their luminosity function
and evolution. Various models of this population can then be
explicitly tested. We cannot yet directly obtain redshifts for mid-IR
sources in the waveband of detection, but must instead use
ground-based optical and/or near-IR observations. Identifications, and
spectra, at these complementary wavelengths must thus be obtained.

We here report the first results from identification and redshift
measurements for a deep 12$\mu$m ISO survey (Clements et al., 1999)
conducted using the ISOCAM instrument (Cesarsky et al. 1996) on the
ISO satellite. The survey covers a total area of 0.1 sq. degrees made
up of four separate subregions separated by several degrees, and
reaches a 3$\sigma$ flux limit of $\sim$300$\mu$Jy. A total of 148
sources were detected above the 3$\sigma$ limit, though a number of
these may still be fragments of brighter sources. Where optical
identifications in the US Naval Observatory Catalogue (USNO-A) are
available, we can achieve star-galaxy separation using an optical-IR
colour-colour method. This relies on the stellar spectra approximating
a single temperature black-body, while the galaxies are assumed to
have a substantial infrared excess. Very few of the galaxies found in
the survey were previously known. Only one object, which was detected
by IRAS, had a measured redshift.  A full description of this 12$\mu$m
survey is given in Clements et al. (1999) (hereinafter Paper 1).  The
goal of the present paper is to present the first results of the
followup programme aimed at this survey.

The rest of this paper is organised as follows. In the next section we
describe the sample of objects observed, and after that the
observations themselves. The results of the spectroscopy are then
discussed followed by brief comments on the identification of 12$\mu$m
sources in two previously blank fields. Discussion of the luminosity
function of the 12$\mu$m sources and then emission line diagnostics
for their power sources follows. We then discuss the nature and
significance of this population before drawing our conclusions. An
H$_0$ of 75kms$^{-1}$Mpc$^{-1}$ and $q_0 = 0.5$ are assumed throughout.

\section{The Sample}

The target list for the spectroscopic followup was constructed from
the list of $>5\sigma$ non-stellar sources given in Paper 1. From
these we then selected those objects with an optical ID brighter than
R=19.6 magnitudes, as cross identified in the USNO-A catalogue. This
provides a list of 25 objects. [The object F3\_15, was found to be a
star in paper 1 on the basis of its optical/IR colours. This stellar
identification was inadvertently missed from Table 2 in that paper,
but was included for the analysis work in that paper. It is thus not
part of the basis sample in the current work.]

As part of the INT Wide Field Survey (McMahon et al., 2000) and the
UKIRT Mini-survey (Davies et al., private communication), much of
Field 1 of Paper 1 was observed in the optical/near-IR, providing
UBVRIJK images and photometry. The identifications and fluxes from
USNO-A can thus be checked and are found to be broadly correct.  The
deeper optical data also allows us to go fainter in our
identifications of ISO sources in Field 1. We find that all of the
$>5\sigma$ ISO sources are identified in this field (the faintest is
F1\_44 which is identified with an R band object with R=23.15). We
also attempt to obtain identifications for fainter $3$ -- 5$\sigma$
significance ISO sources using the new optical data. We find good
identifications by eye for all except three ISO sources $>3\sigma$
within the area covered by the INT WFS image. These three unidentified
sources are all of low signal-to-noise, (3.5, 3.0 and 3.2 $\sigma$)
and so might be residual noise spikes or cosmic hits that have managed
to escaped our ISOCAM reduction. The reliability of the source
detection, discussed in paper 1, with only 4\% of initially detected
sources being rejected as spurious cosmic hits, suggests the
that these may be real sources. We thus conclude that the
optical imaging, to about R=23.5, which identifies 26 ISO sources and
fails on these three, has a maximum identification incompleteness of
$\sim 11\%$. This bodes well for obtaining complete identifications
for the $3$ -- 5$\sigma$ sources in the other three fields with deeper
optical imaging, which has now been obtained (Clements et al., in
preparation).

The basic properties of the sources observed in this project are given in
Table 1.

\section{Observations}

Most of the observations were carried out at the ESO 3.6m telescope at La
Silla from Nov. 13 to 15 1999. We used the EFOSC2 instrument. This
provides both imaging and long slit spectroscopic capabilities, using
filters and grisms (Patat. 1999). The detector is a 2060x2060 thinned
Loral CCD which has good sensitivity throughout the optical
spectrum. The pixel scale is 0.157''/pixel. Most of the spectra
discussed here were taken using a 2'' wide slit, though a few used a
1'' slit, and Grism 12, which covers the spectral region from $\sim$
6000 to 10000 \AA~ at 2.12\AA/pixel dispersion. A few objects were
also observed in the blue using Grism 11 which covers $\sim$3400 to
7500\AA~ at the same resolution. Wavelength calibration was achieved
using a He-Ar arc lamp attached to the spectrograph, while flat
fielding used a dome flat. The contribution of sky lines to the
spectra was calculated by fitting and subtracting a cubic spline
function to the spatial direction for each detector row in the
dispersion direction. Standard IRAF tasks were used for these
procedures. The CCD response function was calibrated out of the
spectra using observations of the spectroscopic standard stars
LTT9239 and LTT1020.
Spectroscopy integration times ranged from 900
to 1800s. Typical observing conditions were for seeing between 1'' and
2'', and with reasonable transparency. The conditions, however, were
not sufficiently stable to be photometric.

Two objects, F1\_0 and F2\_0 were observed on a previous run, from 24
to 25 December 1997, which was severely affected by poor conditions
and equipment problems. This was also conducted at the ESO 3.6m using
EFOSC2.  The observational setup was broadly similar, though a
different grism, Grism 1, was used. This covers $\sim$3200 -- 10000
\AA~ at 6.3\AA/pixel. The observations and data reduction for these
objects followed the same scheme described above. The reduced
resolution, however, means that these observations cannot be used for
emission line diagnostics.

In addition to the spectroscopic observations, R band imaging was also
obtained for two objects during the 1999 run. These were F4\_2, and
F4\_11, chosen to be bright 12$\mu$m sources, with strong detections
in the ISOCAM observations, but without a bright (R$<$19.6) optical
identification. The images were taken as a series of four separate
200s integrations each shifted by -30'' in both RA and DEC. The images
were flat fielded using an imaging dome flat, and them median stacked
with the appropriate shifts taken into account. Residuals left after
the dome flat were them corrected be median stacking the images for
both objects with each other without the shifts corrected, to provide
a secondary, sky flat image.

\begin{table*}
\begin{tabular}{cccccccc} \hline
Object&RA (2000)&Dec (2000)&F$_{12}$ ($\mu$Jy)&R$_{mag}$&Redshift (z)&Log(12$\mu$m Luminosity) (L$_{\odot}$)&Log(Bolometric Luminosity) (L$_{\odot}$)\\ \hline
F1\_0&03 05 36.2&-09 31 21.6&12147$\pm$200&17.1&0.113&10.1&11.2\\
F1\_3&03 05 15.1&-09 31 17.6&3835$\pm$99&18.6&0.191&10.1&11.2\\
F1\_4&03 05 05.3&-09 32 07.7&2150$\pm$185&19.5&0.249&10.1&11.2\\
F1\_5$^a$&03 05 24.5&-09 35 50.7&1046$\pm$102&20.2&0.478&10.6&11.7\\
F1\_7&03 05 39.5&-09 31 25.7&968$\pm$241&19.1&0.114&9.0&10.1\\
F1\_9&03 05 35.9&-09 31 43.8&3732$\pm$142&18.6&0.115&9.6&10.7\\
F1\_10&03 05 06.1&-09 32 41.5&1903$\pm$185&17.9&0.113&9.3&10.5\\
F1\_11&03 05 14.9&-09 31 03.8&1324$\pm$113&18.9&0.290&10.1&10.8\\
F1\_12&03 05 28.4&-09 35 17.5&842$\pm$92&16.9&0.461&10.4&11.1\\
F1\_18&03 05 36.3&-09 31 32.2&2190$\pm$171&18.5$^c$&0.113&9.3&10.5\\
F1\_30&03 05 30.7&-09 33 13.4&636$\pm$88&18.8&0.138&9.0&9.7\\
F1\_34&03 05 39.0&-09 38 32.2&563$\pm$115&18.3&0.098&8.6&9.3\\
F1\_48&03 05 11.0&-09 33 10.0&605$\pm$119&19.1&0.120&8.8&9.9\\
F2\_0&03 01 06.1&-10 44 27.5&10479$\pm$260&13.0&0.032&8.8&9.9\\
F2\_3&03 00 37.3&-10 42 51.0&2385$\pm$100&18.8&0.171&9.8&10.5\\
F2\_24&03 00 35.4&-10 43 22.3&814$\pm$118&18.5&0.119&9.0&9.7\\
F2\_80&03 00 39.5&-10 39 43.4&522$\pm$89&19.5&0.093&8.5&9.6\\
F3\_3&03 10 00.9&-08 39 39.2&1760$\pm$95&14.6&M star\\
F3\_4&03 10 09.1&-08 38 57.2&1682$\pm$105&17.6&0.107&9.1&10.2\\
F3\_5&03 09 44.7&-08 32 55.6&1374$\pm$108&17.0&0.132&9.3&10.4\\
F3\_34&03 09 46.1&-08 32 44.0&639$\pm$105&17.9&0.576&10.6\\
F4\_1&03 04 03.4&-10 01 18.0&3670$\pm$149&17.9&0.175&10.0&11.1\\
F4\_3&03 03 55.2&-09 56 59.9&3207$\pm$84&16.6&0.171&9.9&10.6\\
F4\_5&03 03 47.3&-09 59 18.2&2481$\pm$215&19.0&0.383&10.7&11.4\\
F4\_6&03 04 06.3&-09 53 47.6&3051$\pm$197&16.6&0.214&10.1&11.2\\
F4\_9&03 03 56.8&-09 57 31.3&1234$\pm$95&17.6&0.171&9.5&10.6\\
F4\_12&03 03 54.7&-09 58 45.6&933$\pm$106&19.1&1.2&12.0&12.7\\ \hline
\end{tabular}
\caption{Properties of 12$\mu$m Sources}
Source name is given by catalogue number in appropriate field (F1 to F4),
as in Paper 1. Objects already known to be stars from Paper 1 have been excluded.
R band magnitudes in field 1 come from the INT WFC survey data, others
come from the USNO-A catalogue. Signal-to-noise ratio for the 12$\mu$m
detection can be estimated from the flux error. Bolometric luminosity is calculated
using the conversion from 12$\mu$m to bolometric luminosity determined
by Spinoglio et al., (1993) and the classifications from emission line
diagnostics. A `mixed' class object is treated conservatively as being
an AGN. Where emission line diagnostics are unavailable the object is
treated as a starburst. Notes; a: additional object below nominal R
band magnitude limit; b: source outside region covered by INT WFC
survey, R band magnitude from USNO-A catalogue; c insecure redshift
based on absorption feature. All other redshifts based on emission
lines.
\end{table*}

\section{Results}

\subsection{Spectroscopic Results}

The reduced spectra were all visually inspected and initial
classifications, as stars, emission line galaxies and, in one case, a
possible absorption line galaxy, were made on this basis. The vast
majority of objects turned out to be emission line galaxies. These
spectra were further analysed to extract redshifts, determine emission
line strengths, and to search for broad emission lines. Redshifts and
derived properties are summarised in Table 2. The following method was
used to extract spectral properties: first the brightest pixel in a
given wavelength range is assumed to be H$\alpha$. Then a composite
spectrum, including a background flux constant wrt. wavelength,
H$\alpha$, [NII] and [SII] emission lines, is matched to the spectrum.
The lines are assumed to have Gaussian shapes, with [NII] and [SII]
lines having the same width but different from H$\alpha$. The
strengths of the H$\alpha$, [NII] and [SII] lines are allowed to vary,
but the ratios of the [SII] and [NII] doublets are fixed at their
physical values (Osterbrock, 1989). The assumed redshift is also
allowed to vary. The $\chi ^2$ between the model and the observed
spectrum is minimised using a downhill simplex method (AMOEBA in
IDL). Where a broad line is suspected, an additional model component
consisting of an extra H$\alpha$ with larger initial Gaussian width is
added to see if the $\chi^2$ is significantly reduced. This is found
to be the case in only one object. The results of these fits are given
in table 3, as are the results of the spectroscopy more
generally. Spectra are shown in Fig 1 and, for the one
quasar in the sample, F4\_12, in Fig.2.

Given the redshifts we then calculate the 12$\mu$m luminosities of the
objects in the usual manner.  We include a K-correction to account for
the effects of the complex spectral energy distribution (SED) in the
mid-infrared. The K-correction adopted is a simple linear fit to the
models of Xu et al.  (1998). Over the redshift range of interest (0 --
1.2) the Xu K-corrections can be approximated as $K=1.7 z$, using the
definition of K-correction from Xu et al. (1998). However, this is the
mean K-correction and does not take into account the effects of SED
differences from one object to another. These differences, though, are
smaller in the ISO 12$\mu$m band than in any of the other ISO filters
used for deep ISOCAM surveys (Xu et al. 1998). The uncertainties
produced by this scatter should be less than $\sim$0.2 mag, and thus
should not have a major effect on the luminosity function we derive
from these luminosities.

\subsubsection{Notes on Specific Objects}

{\bf F3\_3:} This object was classified as a galaxy based on its location
on the B-R-F$_{12}$ colour/colour diagram (Paper 1).
However, spectroscopy clearly shows that the object is an M star. The
previous misclassification is probably due to the large molecular absorption
bands in the M-star spectrum suppressing optical emission below what
would be expected from a simple black-body spectrum.
\\~\\
{\bf F1\_34:} The H$\alpha$ emission line in this object appears to be
broad, $\sim$1000 kms$^{-1}$ (cf. 400kms$^{-1}$ for the other, narrow
lines) indicating the likely presence of a Seyfert 1 active nucleus.
\\~\\
{\bf F4\_12:} At z=1.2 this is the highest redshift object in the
sample.  It is also the only object in the sample characterised by a
clear broad line spectrum, containing emission lines of MgII and
AlII/CIII]. There also appears to be a broad emission feature to the
blue of MgII which could be due to FeII. The MgII FWHM is found to be
4300$kms^{-1}$ (observed frame). The spectrum is shown in Figure 2.

\begin{table*}
\begin{tabular}{ccccccc} \hline
Object&Redshift&H$\alpha$ EW&[NII] EW&[SII] EW&H$\alpha$ (broad)&Class\\ \hline
F1\_3&0.191&104.6&60.7&26.4&&HII\\
F1\_4&0.249&41.5&21.0&15.0&&HII\\
F1\_5&0.478&50.3&33.4&$<$48&&?, HII$^*$\\
F1\_7&0.114&14.7&4.9&1.8&&HII\\
F1\_9&0.115&26.7&11.1&10.0&&HII\\
F1\_10&0.113&17.0&6.4&4.3&&HII\\
F1\_11&0.290&21.7&35.9&$<$3.0&&M, HII$^*$\\
F1\_12&0.461&54.9&53.0&30.7&&SII, SII$^*$\\
F1\_18&0.113&79.1&50.1&24.9&&HII\\
F1\_30&0.138&28.1&20.1&2.7&&M\\
F1\_34&0.098&9.1&13.6&3.4&26.4&S1\\
F1\_48&0.120&61.3&26.7&21.2&&HII\\
F2\_3&0.171&13.25&13.3&8.8&&SII\\
F2\_24&0.119&17.3&20.2&17.2&&SII\\
F2\_80&0.093&56.9&17.1&18.1&&HII\\
F3\_4&0.107&49.6&21.0&13.2&&HII\\
F3\_5&0.132&21.2&10.9&A band&&HII\\
f4\_1&0.175&35.0&20.9&10.3&&HII\\
F4\_3&0.171&14.1&11.9&5.7&&SII\\
F4\_5&0.383&46.7&44.8&9.5&&M, SII$^*$\\
F4\_6&0.214&11.3&6.6&1.8&&HII\\
F4\_9&0.171&A band&A band&&?\\
F4\_12&1.2&&&&&SI\\
\end{tabular}
\caption{Emission Line diagnostics and Classifications}
Emission line equivalent widths and derived classifications for
objects where emission line classifications are possible. HII
indicates an HII region--like spectrum, and thus an object likely to
be powered by a starburst. SI and SII indicate a Seyfert 1 (ie. broad
line) and Seyfert 2 spectrum respectively, and thus the presence of,
and probable powering by, an AGN. M indicates a spectrum that may be a
mixture of HII and Sy2, or may be a LINER -- the current limited set
of emission lines cannot discriminate between these possibilities. 'A
band' indicates that the relevant lines are affected by atmospheric
absorption by the A band, leading to uncertain equivalent widths. In
most of these cases classification is not possible. '?' indicates that
the current measurements or limits to equivalent widths do not permit
classification. In some cases there are second classifications indicated by
$^*$. These are classifications which include the H$\beta$ and [OIII]
fluxes when they are available. In the case of M and ? classifications these
later, more complete, classifications are adopted.
\end{table*}

\begin{figure*}
\vspace{5cm}
\caption{Observed Spectra}
Spectra for the galaxies observed in this programme. The locations of
important features, typically H$\alpha$ and SII, are indicated, and
the wavelength range shown is chosen to highlight these features.
Spectrophotometric callibration is not fully calibrated so flux values are
indicative only.
The objects F1\_0 and F2\_0 were observed in poor conditions on an
earlier observing run. These two spectra are not flux calibrated, merely
divided through by a standard star, so the apparently blue continuum slope
is not real.
\end{figure*}

\begin{figure*}
\psfig{file=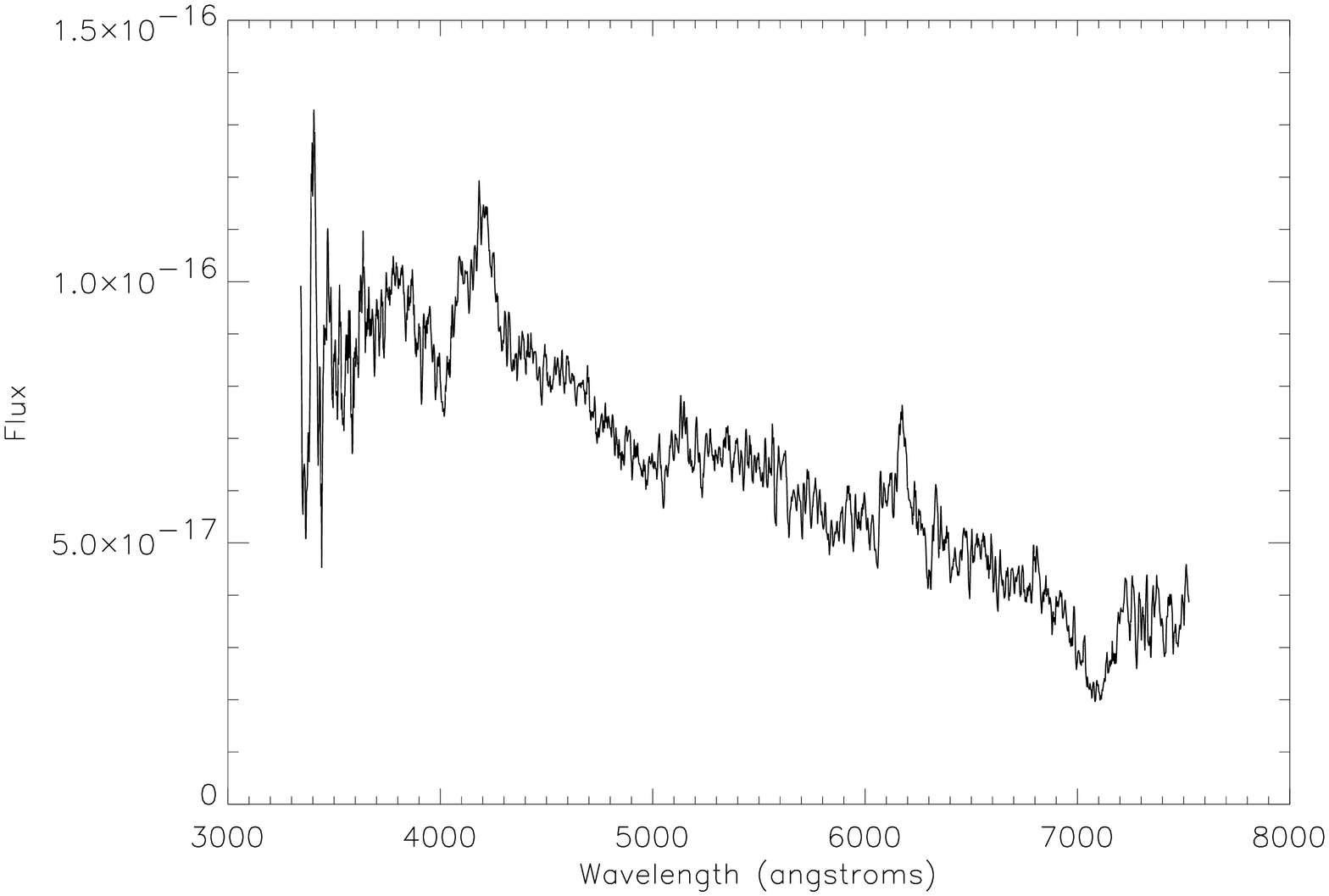,width=12cm,angle=+90}
\caption{F4\_12 Spectrum}
Spectrum for the 12$\mu$m selected quasar discovered in this survey.
\end{figure*}

\section{Blank Fields}

Ten sources from Paper 1 are as yet optically
unidentified. At the same time as our optical spectroscopy, we thus
also obtained moderately deep images for two 12$\mu$m optical blank
field sources, F4\_11 and F4\_2. These are shown in Figure 3. In both
cases we find plausible faint optical identifications within 7'' of
the ISO source position. In the case of F4\_11 the identification is
isolated and unambiguous. In the case of F4\_2, however, there would
appear to be a group of similar faint objects close to the primary
identification. There is thus the potential that the ISO source may be
a combination of two, possibly interacting, objects, or that the
identification might be one of the other optical sources in the field.

We can obtain an estimate for their magnitude by comparison with
magnitudes of known objects imaged at the same time. We find that
F4\_11 has R$\sim$20 and F4\_2 R$\sim$22. These results are consistent
with the results from the INT WF survey data discussed above.

With identifications obtained for these two sources, there remain 8
sources with ISO detections $>5\sigma$ that still lack optical
identifications. Since these are the objects with the highest 12$\mu$m
to optical ratios, they are clearly interesting targets. Their optical
faintness suggests that they are at high redshifts, and they are thus
likely to be highly luminous. Further observations will be proposed to
obtain identifications and redshifts for these, and fainter ISO
sources.

\begin{figure*}
\begin{tabular}{cc}
\psfig{file=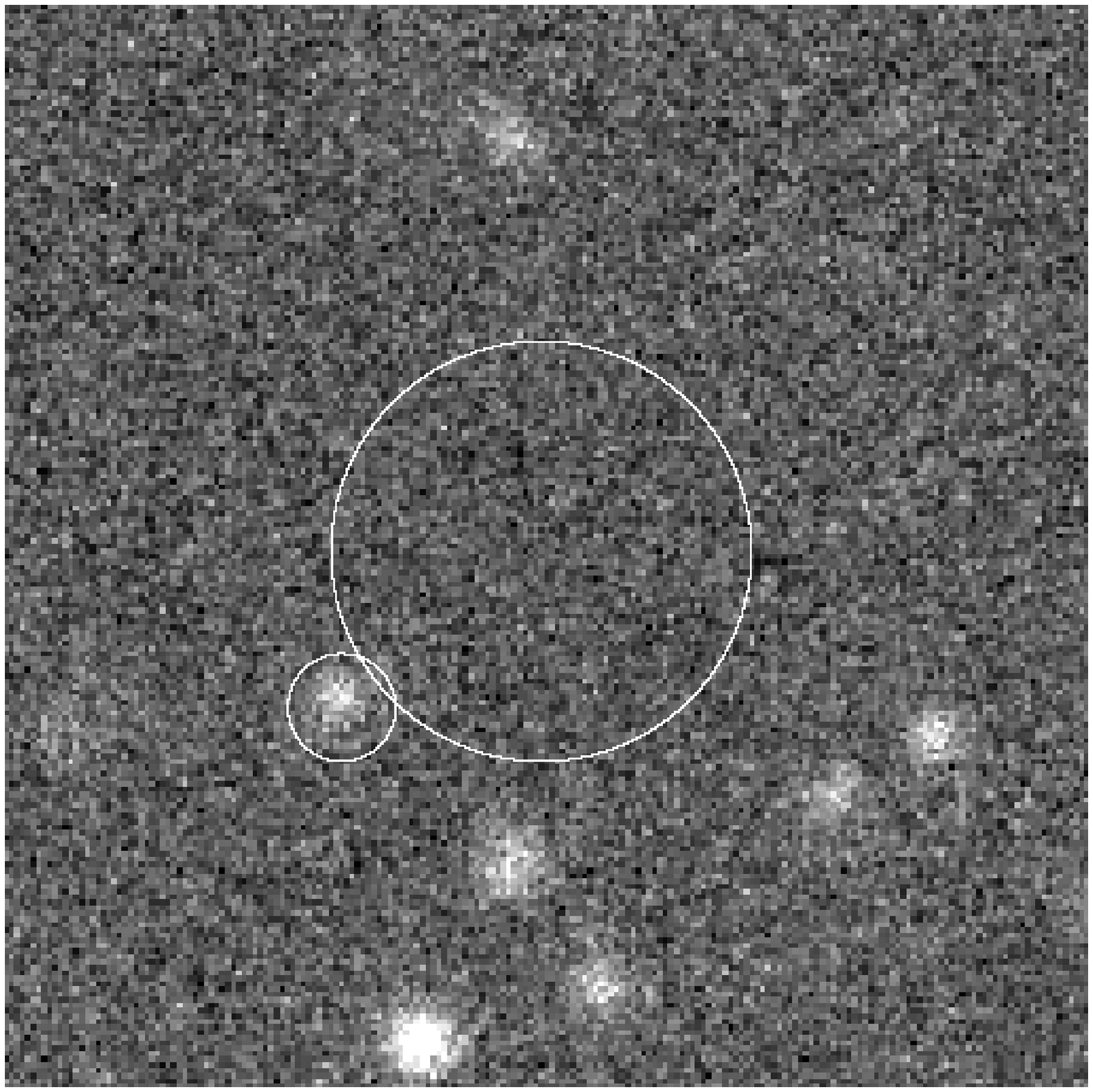,width=8cm}
\psfig{file=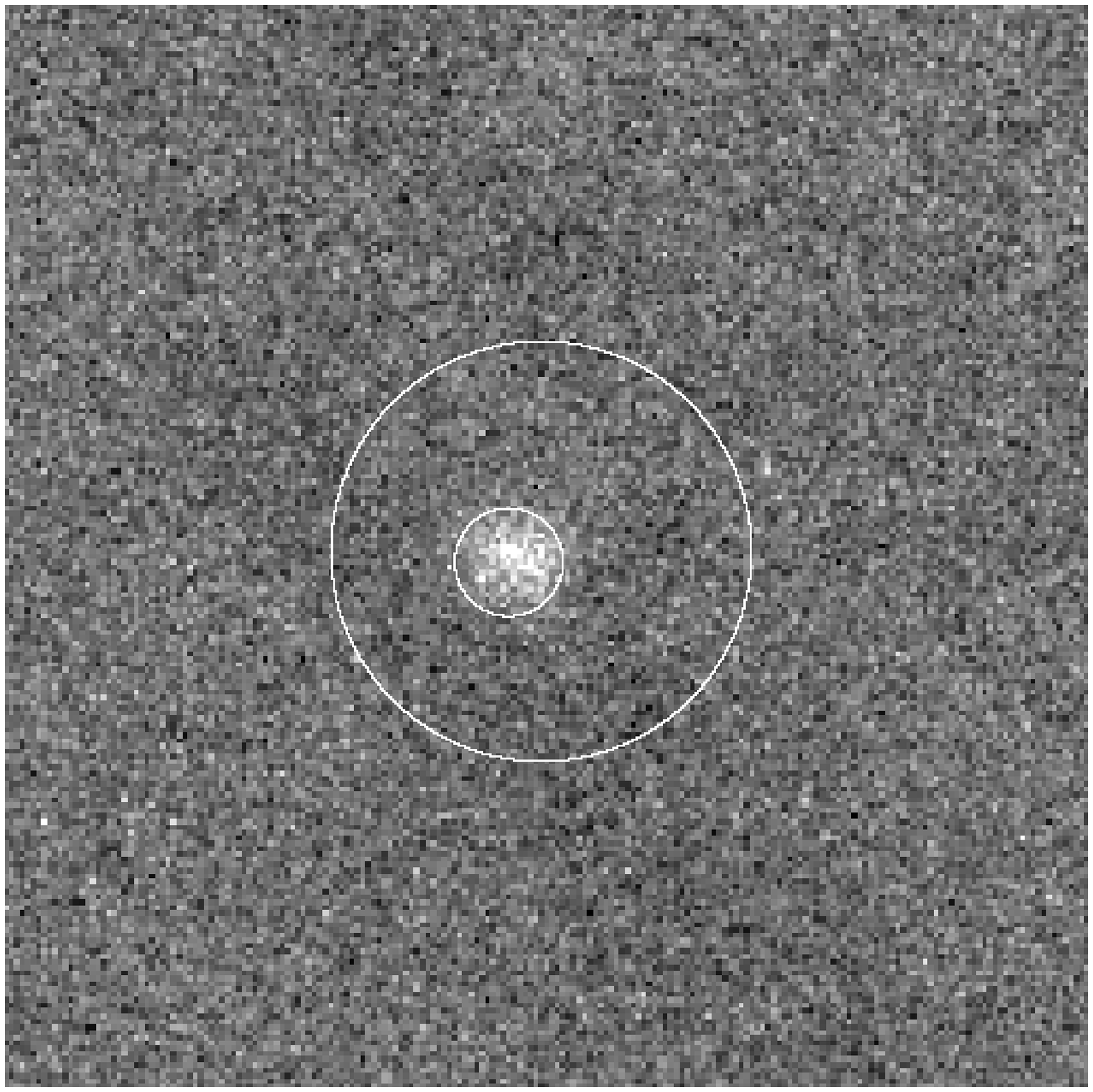,width=8cm}\\
\end{tabular}
\caption{R band images of 12$\mu$m optical blank fields.}
F4\_2 (left) and  F4\_11 (right). The images are 31.4'' to a side. The nominal
position of the ISO source is at the centre of each image. The astrometric
accuracy of the ISO observations is 6'' (2$\sigma$) (Paper 1), and this is
shown by the large circle. A smaller circle highlights the likely
identification. The F4\_2 identification has R$\sim$22, F4\_11 has R$\sim$20. See text for
details. Images are of similar depths, reaching R$\sim$23. The scaling is linear,
chosen to highlight relevant features.
\end{figure*}

\section{The Luminosity Function}

The data presented in this paper provides redshift information for a
complete subsample of the 12$\mu$m survey. We are complete to the
5$\sigma$ flux limit of the survey, and to R=19.6 magnitudes for the
optical identifications and redshift information. We can thus use the
{\em accessible volume} method (Avni \& Bahcall, 1980) to determine
the 12$\mu$m luminosity function in this sample, whose 12$\mu$m flux
limit is a factor of $>$ 20 deeper than existing determinations of the
mid-IR luminosity function (Hacking et al., 1987). We use the standard
1/V$_{max}$ method to determine the luminosity function, with the
volume determined from the maximum distance at which a source would be
detected ie. the accessible volume. The luminosity function
contribution of each object is 1/V$_{max}$ where V$_{max}$ is given by
the redshift at which an object would drop out of either the survey
data or the followup programme, whichever is smaller.

Thus:

\begin{equation}
\Phi (L) dL = \sum_{i} \frac{1}{V_i}
\end{equation}

where $\Phi(L)$ is the number of objects $Mpc^{-1}$ in the luminosity range
$L$ to $L+dL$ and

\begin{equation}
V_i = min( V_{opt,i}, V_{12,i} )
\end{equation}

where $V_{opt,i}$ is the volume in which the $i$th object could have
been detected down to the R=19.6 magnitude limit of the optical
followup, and $V_{12,i}$ is the volume in which the $i$th object could
have been detected down to the 500$\mu$Jy flux limit in the original
survey.  The volumes are calculated taking into account the variable
flux depths of the original 12$\mu$m survey (see Paper 1). 

The luminosity function produced from these considerations is shown in
figure 4.  As can be seen the luminosity function is a fairly close
match to existing 12$\mu$m luminosity function determinations based on
low redshift IRAS data (see eg. Fang et al 1998). The only substantial
excursion from the local luminosity function occurs at the highest
luminosities. Since these high luminosity bins are also dominated by
the highest redshift objects this might be taken as evidence for
evolution. Combining the two highest luminosity bins to provide better
statistics (3 objects) at the highest luminosities suggests a $\sim 3
\sigma$ significance difference between the current luminosity
function and that derived from IRAS (Fang et al. 1998; Rush et al.,
1993).  The mean redshift for objects in this bin is 0.7, implying
density evolution at a rate of $(1+z)^8$ or luminosity evolution at a
rate of $(1+z)^5$ with large uncertainties on these exponents. These
values are broadly consistent with the evolution in the ultraluminous
IRAS galaxies suggested by Kim \& Sanders (1998) and with that
suggested by Xu (2000) for the 15$\mu$m population.  However, we as yet
have only a small number of objects at the highest
luminosities. Indeed, the highest redshift bin contains only one
object, F4\_12. This is one of only two broad-line objects in the
survey as well as being at the highest redshift, so might be an
unusual object biasing the statistics. There may also be additional biases
entering the data, eg. Malmquist or Eddington bias, since the sample size
is so small. These issues will be addressed when the identification
programme is complete and we have a larger number of sources.

These hints at evolution in the high luminosity/redshift end of the
population can only be confirmed by determining redshifts for the rest
of the sample. Since the remaining galaxies are optically faint
(R$>$19.6) there is a strong possibility that substantially more high
redshift galaxies will be found, allowing a much better examination of
the high redshift and high luminosity end of the luminosity function.

\begin{figure*}
\psfig{file=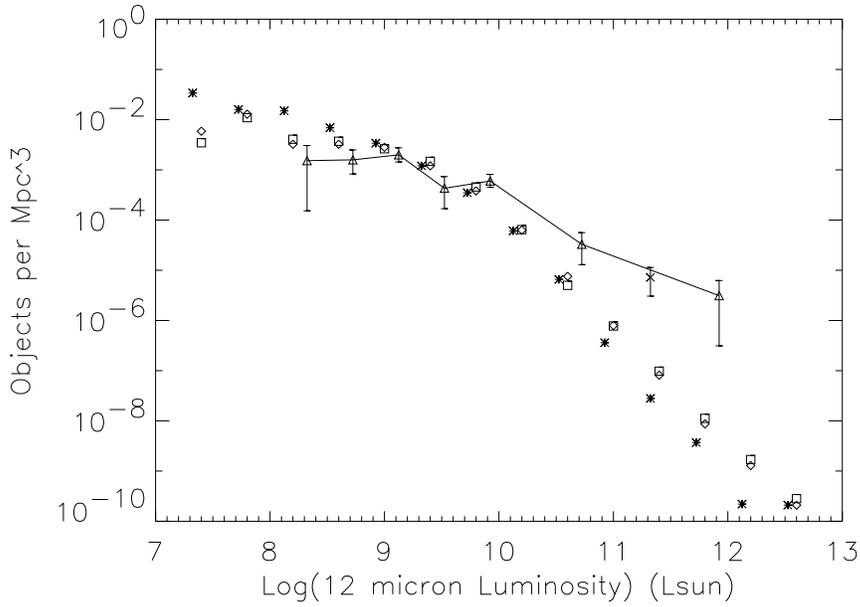,width=12cm,angle=+90}
\caption{The 12$\mu$m Luminosity Function}
The 12$\mu$m luminosity function from Fang et al. 1998 (open squares and
diamond for two different corrections to the local density field),
Rush et al. 1995 (with luminosity divided by a factor of 1.5 to correct for
photometric problems from the use of ADDSCAN data (Alexander \&
Aussel, private communication)) and from the present work (triangles
joined by a line). Note the similarities between the older, IRAS LF
determination and the present, ISO determination, and the possible
departure at high luminosities which equate to high redshifts. Also
shown (X) are the last two bins of the new luminosity function combined
to provide a better determination of the evolution at the highest redshifts.
\end{figure*}

\section{Emission Line Diagnostics}

The nature of the ionising spectrum in a galaxy can be determined by
examining emission line ratios (Baldwin et al., 1981). Standard ratios
include H$\alpha$/[NII] and H$\alpha$/[SII], which can be determined
for most of the objects in the current sample, as well as
[OI]/H$\alpha$, [OIII]/H$\beta$ and [OII]/H$\beta$. These latter
ratios are not available for most of the objects in the sample since
our spectral range did not extend far enough into the blue, or go deep
enough to detect the weak [OI]6300 line. Consideration of the standard
emission line diagnostic plots shows that we can attempt a
classification using just the H$\alpha$, [NII] and [SII] lines we have
for the majority of the objects. Our classification diagram is shown
in Figure 5. On this basis we find that the majority of our 12$\mu$m
sources (11/16 where appropriate data is available) have HII
region-like spectra ie. are powered by starbursts, while 4/16 have
Seyfert2-like spectra ie. contain and may be powered by active
nuclei. Two further objects have broad line Seyfert1-like spectra,
producing a total of 6/18 objects that contain AGN.  Three objects are
not clearly classified by this technique and could be composite
objects or LINERs.

We can extend this classification scheme for the
four objects at high enough redshift and with sufficient
signal-to-noise in their spectra to detect H$\beta$ and [OIII]. Of
these objects only one is fully classified using just
H$\alpha$, [NII] and [SII]. This object, F1\_12, is classified as a
Sy2 by both methods. Of the other three objects where we can use
H$\beta$ and [OIII], F1\_5 was previously unclassified due to a poor
[SII] detection, and the others, F1\_11 and F4\_5, have mixed
classifications. Once H$\beta$ and [OIII] ratios are included we find
that F1\_5 and F1\_11 appear to be starbursts (ie. HII-region-like).
F4\_5 appears to have a confused classification, with [OIII]/H$\beta$
vs. [SII]/H$\alpha$ indicating an HII region spectrum, and
[OIII]/H$\beta$ vs. [NII]/H$\alpha$ indicating a Sy2 spectrum. On
further examination it appears that the [SII] line strength may be
underestimated due to a sky absorption feature at the extreme red of
the optical band ($\sim$9300\AA). We thus tentatively conclude that
it is in fact a Sy2.

The final tally is thus 14 starbursts, 7 AGN (Sy1 and Sy2), 1 mixed
and one unclassifiable. Of the objects where we have a firm
classification, 1/3 (7/21) are AGN-type and 2/3 (14/21) are
starbursts.  We find no significant difference between 12$\mu$m
luminosities of the two classes, HII and Seyfert, though the most
luminous objects are from the Seyfert class.

The Rush et al. (1993) sample of local 12$\mu$m selected objects had
an AGN fraction of 13\%, so the larger fraction we have found here,
33\%, may represent some evolution in the population. Alternative
explainations, such as different K-corrections between AGN and non-AGN
containing objects, greater prominance in the optical for AGN objects,
are also possible. These possibilities will be investigated when the
identification programme for the complete 12$\mu$m survey is finished.

\begin{figure*}
\begin{tabular}{cc}
\psfig{file=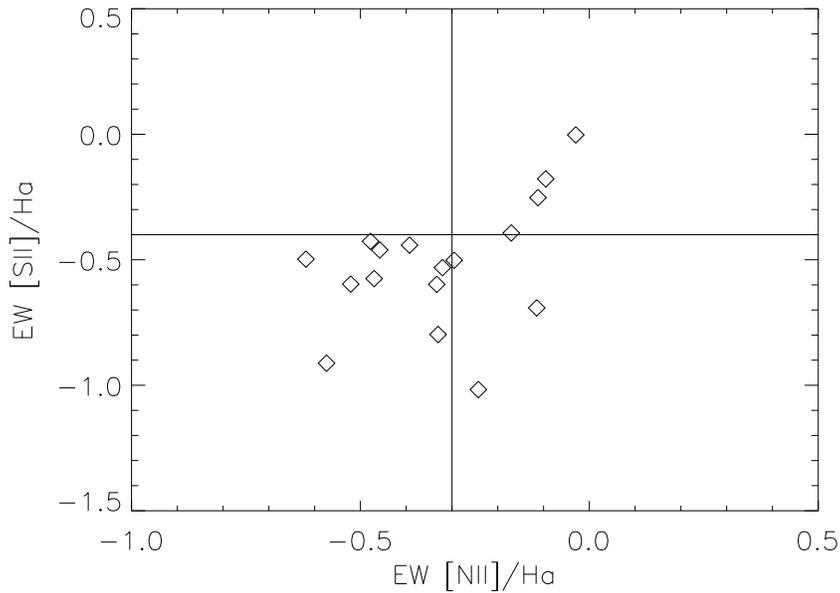,width=12cm,angle=+90}
\end{tabular}
\caption{Emission Line Diagnostics}
Starburst-type objects will lie in the lower left corner, AGNs in the
upper right. Objects in the lower right corner receive a mixed
classification, and may be composite objects or possibly LINERs. The
broad line objects, F4\_12 and F1\_34, are not plotted on this
diagram.
\end{figure*}

\begin{figure*}
\psfig{file=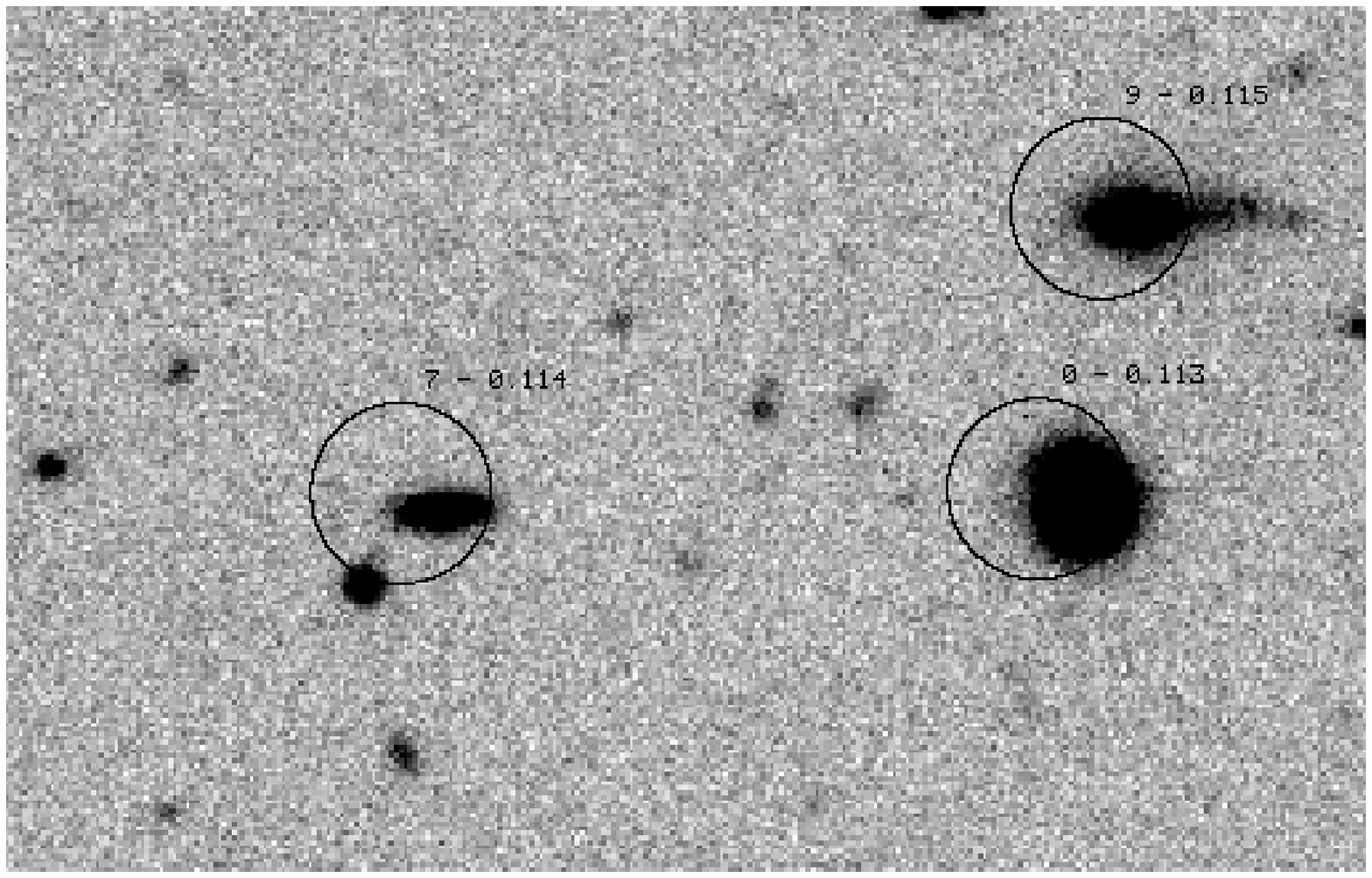,width=12cm,angle=-90}
\label{group}
\caption{Group of Objects in Field 1}
Circles show the 6 arcsecond 2$\sigma$ astrometric errors on the ISO
source position. Text indicates the number of the source in the field
1 list and its redshift. Image is 100 by 66 arcseconds. Note also the
disturbed morphology of source F1\_9. North is up and east is to the left.
Scaling is linear, and chosen to highlight relevent features.
\end{figure*}

\section{Discussion}

\subsection{Physical Associations}

If we examine the redshift distribution, we find several clumps of
objects that appear to group in redshift. These include an apparent group
around z=0.11 in field 1, and around z=0.17 in field 4. We can test
the reality or otherwise of these associations by examining the
positions of the objects on the sky. In field 1 we do indeed find that
the objects F1\_0, F1\_7 and F1\_9 appear to have a close physical
association. They form a small group covering a region $\sim$ 1
arcminute across, corresponding to $\sim$ 100kpc at z=0.11,
using our assumed cosmology (see Fig. 6). These objects may thus
be part of a small group or weak cluster. One other object in field 1
is at roughly the same redshift, F1\_10 at z=0.113. It is much further
away on the sky, 7.5 arcminutes, corresponding to 750kpc. It might
thus be part of some weak larger scale structure, but this is
difficult to assess with the small numbers of redshifts we have. A
simple analysis, using the luminosity function derived in section 6,
suggests that the association of F1\_10 with the other three objects
is only marginally significant, at $\sim$90\% level.

In field 4 we find that F4\_9 and F4\_3 are close in both redshift
(difference $<$0.0005) and on the sky (39 arcsecond), with a separation
corresponding to 94kpc at z=0.17. They are thus likely to be a
close pair of galaxies. F4\_1, in contrast, is significantly
further away in both redshift (0.004 in z) and angular separation
(4.8 arcminutes). It is thus unlikely that there is a strong physical
association between F4\_1 and the other two objects. 

It is thus possible that 5 of our 12$\mu$m sources appear in
two groups, though better statistics are needed to determine if the
correlation function of 12$\mu$m selected objects is any different from
optically selected, normal galaxies.

\subsection{Bolometric Luminosities}

Spinoglio et al. (1995) found a correlation between 12$\mu$m
luminosity and bolometric luminosity. The basis for this correlation
was argued to be that the wavelength is a `pivot point' between the
optical emission, which rises more slowly than linearly with
bolometric luminosity, and the mid- to far-IR, which rises more
rapidly than linearly with bolometric luminosity. They also found that
the constant of proportionality relating 12$\mu$m and bolometric
luminosity is different for starburst and AGN sources --- thus
$L_{bol} = 14 L_{12}$ for starbursts and $L_{bol} = 5 L_{12}$ for
AGNe, where $L_{12} = \nu(12 \mu m) L_{\nu}(12\mu m)$ in solar
units. If we assume that our optical classifications of these objects
genuinely identify the source of their luminosity, then we can
calculate the bolometric luminosity of the sources. These are given in
Table 2.  We find that the typical bolometric luminosity for our
12$\mu$m sources is around 10$^{10.5} L_{\odot}$. The population of
mid-IR sources so far uncovered would thus appear to have relatively
moderate luminosities, and to not represent any substantial new
population of dust-enshrouded starbursts or AGN. Our highest
luminosity source, F4\_12, however, turns out to have a bolometric
luminosity of 10$^{12.7} L_{\odot}$, approaching that of Hyperluminous
infrared galaxies. Since this is the highest redshift object in the
sample, and our luminosity function analysis seems to indicate strong
evolution at high redshifts or high luminosities, then one can
speculate that this might be the first member of a lurking population
of obscured high luminosity systems that we have yet to uncover.
Completion of our identification programme can test this idea.

\subsection{The Nature and Evolution of the 12$\mu$m MilliJansky Population}

Low luminosity 12$\mu$m selected sources appear to be similar to the
local samples. We find that about 1/3 of them contain AGNe while the
remainder are powered by starbursts. These fractions are somewhat
higher than those discussed by Rush et al. (1993). At the highest
luminosities there are hints of strong evolution being detected, but
these are currently dominated by small number statistics. With deeper
observations we will not only be able to confirm the presence of
evolution but also test claims based upon the 15$\mu$m surveys that a
substantial new population of objects becomes significant at the
$\sim$1mJy flux level (eg. Elbaz et al., 1998; Aussel et al.,
1999). It should be noted that selection of objects at 12$\mu$m is
much less affected by K-corrections due to the complex mid-IR spectral
energy distributions of star-forming galaxies and AGN (Xu et al.,
1998) than selection at 15$\mu$m. It has been suggested (Xu, 2000)
that the bump in 15$\mu$m number counts at $\sim$ 0.4mJy is entirely
due to this K-correction effect and not due to any new
population. While the current survey is working at somewhat brighter
fluxes we find no clear evidence for such a population. However, if
such sources are optically faint, whether intrinsically or as a result
of their redshift, they may well have been missed by the optical
selection criterion. Completion of the optical identifications and
redshift measurements for all the objects in the survey is needed to
test these claims and is thus a high priority.

\subsection{Future Prospects}

The present paper is only the first step in following up the galaxies
found in this 12$\mu$m ISO survey. We have here discussed the results
of redshift measurements on 26 of the 36 galaxies with fluxes above
the 5$\sigma$ flux limit of our survey, forming a complete subsample
with R$<$19.6. While we are able to obtain interesting conclusions,
they are clearly limited by small number statistics, and we are as yet
unable to study objects with a large F$_{12}$/F$_{opt}$ ratio
(eg. F4\_2 which has F$_{12}$ of 3.6mJy, but an R mag. of $\sim$22).
Completing the redshift observations for the 10 $> 5\sigma$ sources so
far unobserved is clearly a high priority.  In field 1, where we have
deeper optical data, nearly all of the 12$\mu$m sources are identified with
objects brighter than R=23.2. If this is representative of the rest of
the sample, then complete identifications and spectroscopy should be
fairly easy to obtain. Perhaps more interesting is the prospect of
extending our redshift observations to fainter 12$\mu$m fluxes. If we
go to a 4$\sigma$ detection limit, rather than 5$\sigma$ we would
include $\sim$ 45 more sources. All of these objects, again, are
detected in our deep optical observations of Field 1, so such a study
is eminently feasible, given deeper optical imaging and a moderate
amount of time for spectroscopy. We will thus be able to substantially
increase the number of redshifts we have for faint 12$\mu$m sources,
and will be able to place the tentative conclusions of the present
paper on a much firmer footing. Such observations will make an
excellent parallel study to the 15$\mu$m surveys also
underway (eg. Oliver et al. (2000); Elbaz et al. (1999)).

\section{Conclusions}
We have obtained spectroscopy for a complete R$<$19.6 subsample of our
12$\mu$m survey with the ISO satellite. The redshifts of these objects
range from 0.032 to 1.2, with derived bolometric luminosities ranging
from 10$^{9.3}$ to 10$^{12.7}$ L$_{\odot}$. There is tentative
evidence for strong evolution at the high luminosity and/or high
redshift end of the luminosity function. We find that 1/3 of these
sources contain an AGN, while the rest seem to be powered by
starbursts. At this stage our conclusions are limited by small number
statistics, but there are very good prospects for extending the
identification programme for this survey, and thereby allowing
stronger, more definitive conclusions to be drawn.  ~\\~\\ {\bf
Acknowledgements}

It is a pleasure to thank Scott Croom and Steve Warren for the
provision of the INT Wide Field Survey optical data for Field 1, and
the UKIRT minisurvey team for providing similar infrared data. The
Digitised Sky Survey was produced at the STSCI under US Government
Grant NAG W--2166. This research has made use of the NASA/IPAC
Extragalactic Database (NED) which is operated by the Jet Propulsion
Laboratory, California Institute of Technology, under contract with
the National Aeronautics and Space Administration. The USNO-A survey
was of considerable help, and we would like to express our thanks to
all those who helped to produce it. We would also like to thank Herve
Aussel, Dave Alexander and Matt Malkan for useful discussions. DLC
acknowledges support from a PPARC postdoctoral position and, during
the early stages of this work, an EU TMR Network post
(FMRX-CT96-0068).  We would also like to thank the anonymous referee
for many useful comments.

\end{document}